\documentclass{article}
\usepackage{spconf,amsmath,graphicx}
\usepackage{cite}
\usepackage{amsmath,amssymb,amsfonts}
\usepackage{algorithmic}
\usepackage{graphicx}
\usepackage{siunitx}
\usepackage{textcomp}
\usepackage{svg}
\usepackage[export]{adjustbox}
\usepackage{xcolor}
\usepackage{bm}
\usepackage{multirow}
\usepackage[caption=false, font=footnotesize, position=top]{subfig}

\ninept
\newcommand{\sidx}{i} 
\newcommand{\cidx}{j} 
\newcommand{\etal}{\textit{et~al.} }

\usepackage[firstpage=true]{background}
\usepackage{hyperref}

\SetBgContents{%
	\fbox{%
	\parbox{\textwidth}{%
		\footnotesize Copyright 2023 IEEE. Published in ICASSP 2023 – 2023 IEEE International Conference on Acoustics, Speech and Signal Processing (ICASSP), scheduled for 4-9 June 2023 in Rhodes Island, Greece. Personal use of this material is permitted. However, permission to reprint/republish this material for advertising or promotional purposes or for creating new collective works for resale or redistribution to servers or lists, or to reuse any copyrighted component of this work in other works, must be obtained from the IEEE. Contact: Manager, Copyrights and Permissions / IEEE Service Center / 445 Hoes Lane / P.O. Box 1331 / Piscataway, NJ 08855-1331, USA. Telephone: + Intl. 908-562-3966.}}%
}	
\SetBgScale{1}
\SetBgAngle{0}
\SetBgPosition{current page.north}
\SetBgVshift{-1.6cm}
\SetBgColor{black}
\SetBgOpacity{1}

\title{A novel Cross-Component Context Model for End-to-End\\Wavelet Image Coding}
\name{Anna Meyer and André Kaup \thanks{The authors gratefully acknowledge that this work has been funded by the Deutsche Forschungsgemeinschaft (DFG, German Research Foundation) under project number 461649014.}}
\address{\textit{Multimedia Communications and Signal Processing} \\
	\textit{Friedrich-Alexander-Universität Erlangen-Nürnberg (FAU)}\\
	Erlangen, Germany \\
	\{ anna.meyer, andre.kaup \}@fau.de}
\begin{document}
%
\maketitle
\begin{abstract}
In contrast to traditional compression techniques performing linear transforms, the latent space of popular compressive autoencoders is obtained from a learned nonlinear mapping and hard to interpret. In this paper, we explore a promising alternative approach for neural compression, with an autoencoder whose latent space represents a nonlinear wavelet decomposition. Previous work has shown that neural wavelet image coding can outperform HEVC. However, the approach codes color components independently, thereby ignoring inter-component dependencies. Hence, we propose a novel cross-component context model (CCM). With CCM, the entropy model for the chroma latent space can be conditioned on previously coded components exploiting correlations in the learned wavelet space. The proposed CCM outperforms the baseline model with average Bj{\o}ntegaard delta rate savings of 2.6~\% and 1.6~\% for the Kodak and Tecnick image sets. Also, our method is competitive with VVC and learning-based methods. 
\end{abstract}
\vspace{-2mm}
\section{Introduction}
\vspace{-2mm}
\label{sec:intro}
\noindent
Neural image compression is a highly active field of research, with most state-of-the-art approaches based on compressive autoencoders \cite{minnen2018joint, Balle2016}. Current work \cite{Guo2020, Cheng2020, Qian2022} has often focused on advancing the autoregressive context model and hyperprior framework proposed by Minnen, Ball\'{e}, and Toderici \cite{minnen2018joint}. Here, the encoder performs a nonlinear learned transform implemented via neural networks, from which the latent representation is obtained. The discretized latents are losslessly compressed by an entropy coder with an adaptive entropy model. Its entropy parameters are predicted based on an autoregressive context model and a hyperprior transmitted as side information. The latent space of such compressive autoencoders consists of typically more than hundred convolutional feature maps with reduced spatial resolution. 
If there is no redundant information, the channels of the latent space are conditionally independent given the hyperprior \cite{Minnen2020}. It has been shown that redundancies between the latent space channels can be exploited \cite{Minnen2020, Guo2020}, indicating that current architectures and training strategies fail to reach this optimum state. It is unclear how remaining redundancies can be modelled.
\begin{figure}[tb]
		\vspace{-3mm}
	\captionsetup[subfigure]{labelformat=empty, justification=raggedright}
	\centering
	\subfloat[\footnotesize Y component]{%
		\hspace{-3.5ex} 	\includegraphics[width=0.154\textwidth]{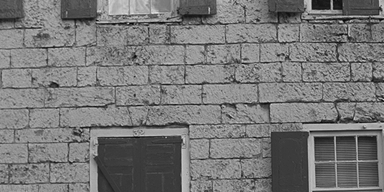}	  
	}
	\subfloat[\footnotesize  Cb component]{%
		\includegraphics[width=0.154\textwidth]{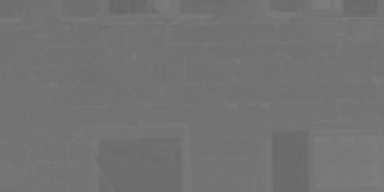}	  
	} \qquad\qquad\qquad\qquad\qquad\quad  \\ 	 \vspace{-9.4ex} 
	\subfloat{%
		\includegraphics[width=0.154\textwidth]{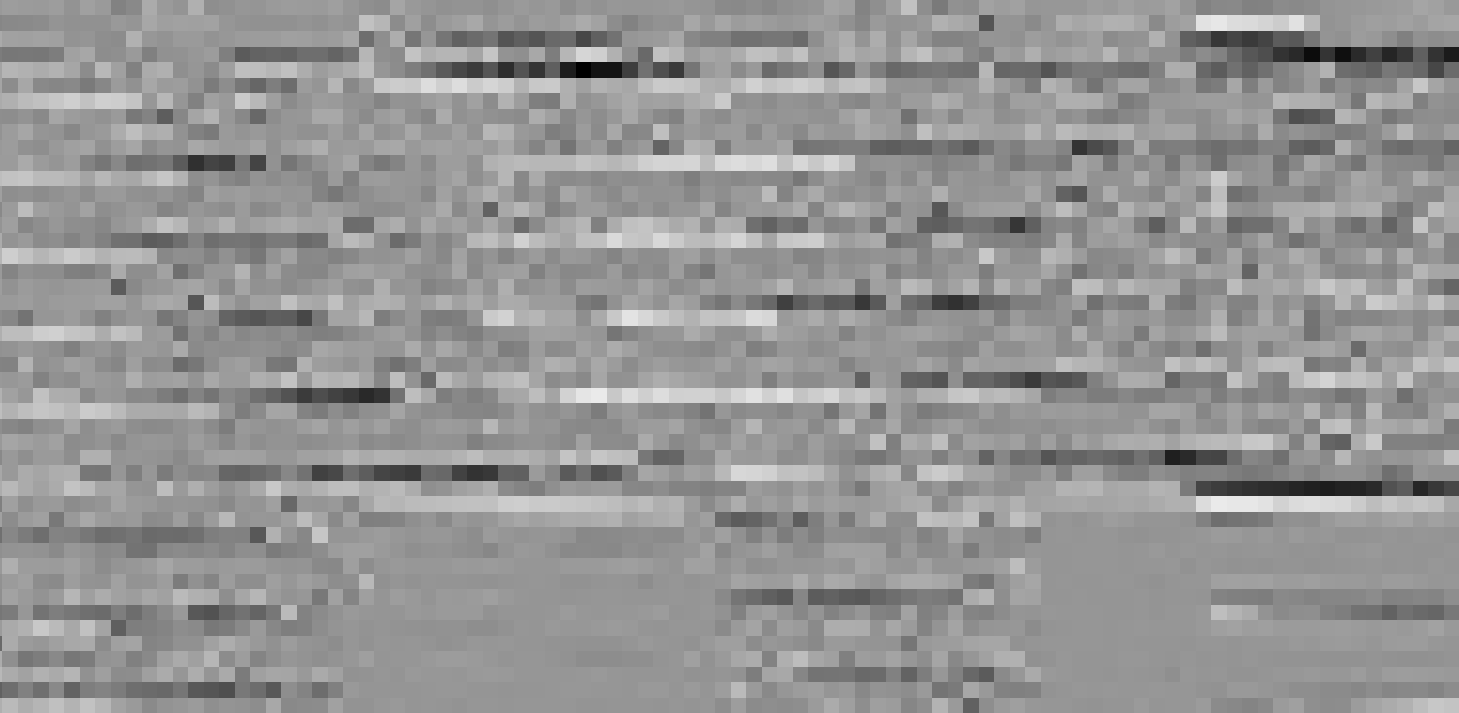}	  	  
	}	
	\subfloat{%
		\includegraphics[width=0.154\textwidth]{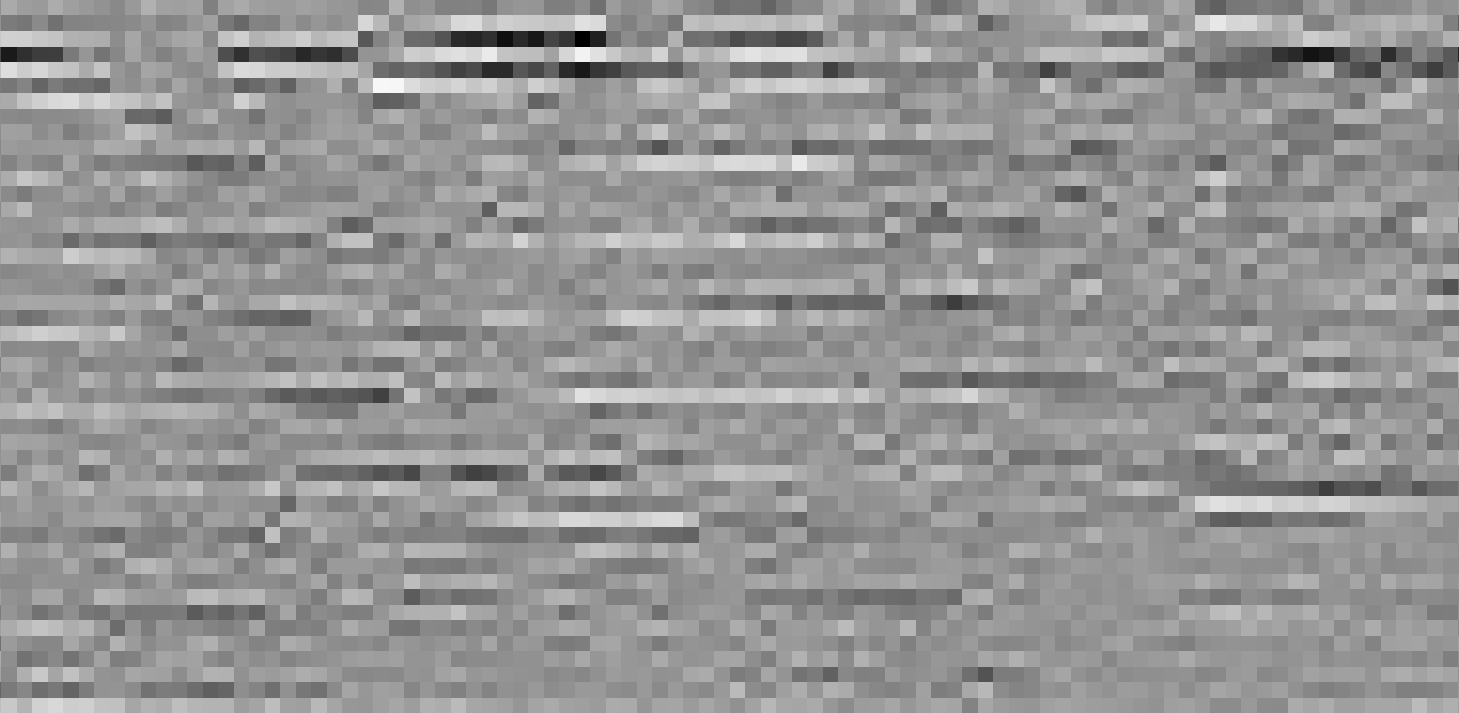}	  
	} 
	\subfloat[\footnotesize Correlation map of Y and Cb subbands HL$_{2}$ and HL$_{3}$]{%
		\includegraphics[width=0.159\textwidth]{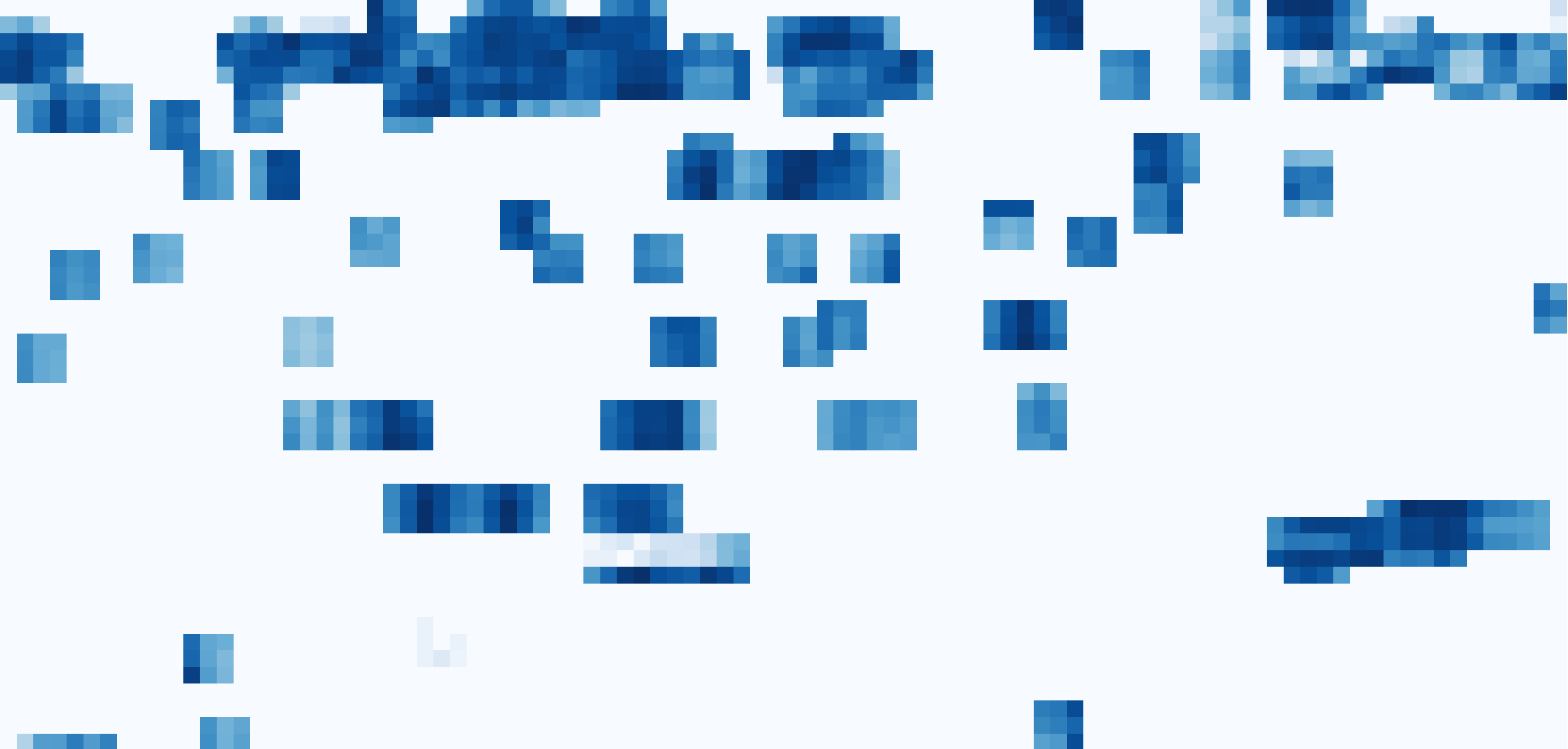}	  
	}
	\subfloat{%
		\hspace*{-1.2ex} \includegraphics[width=0.017\textwidth]{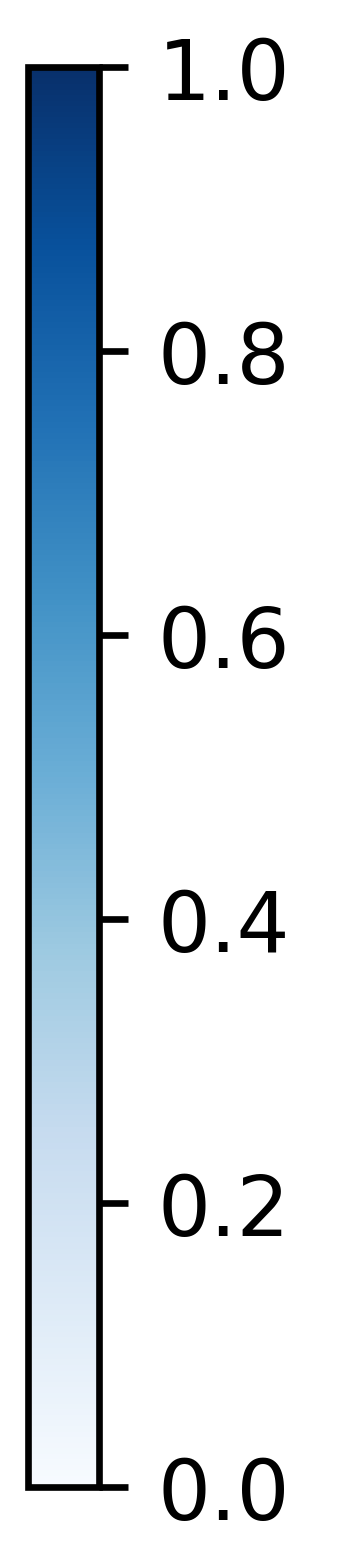}
	} 
	\\ \vspace{-1.9ex}
	\subfloat{%
		\includegraphics[width=0.154\textwidth]{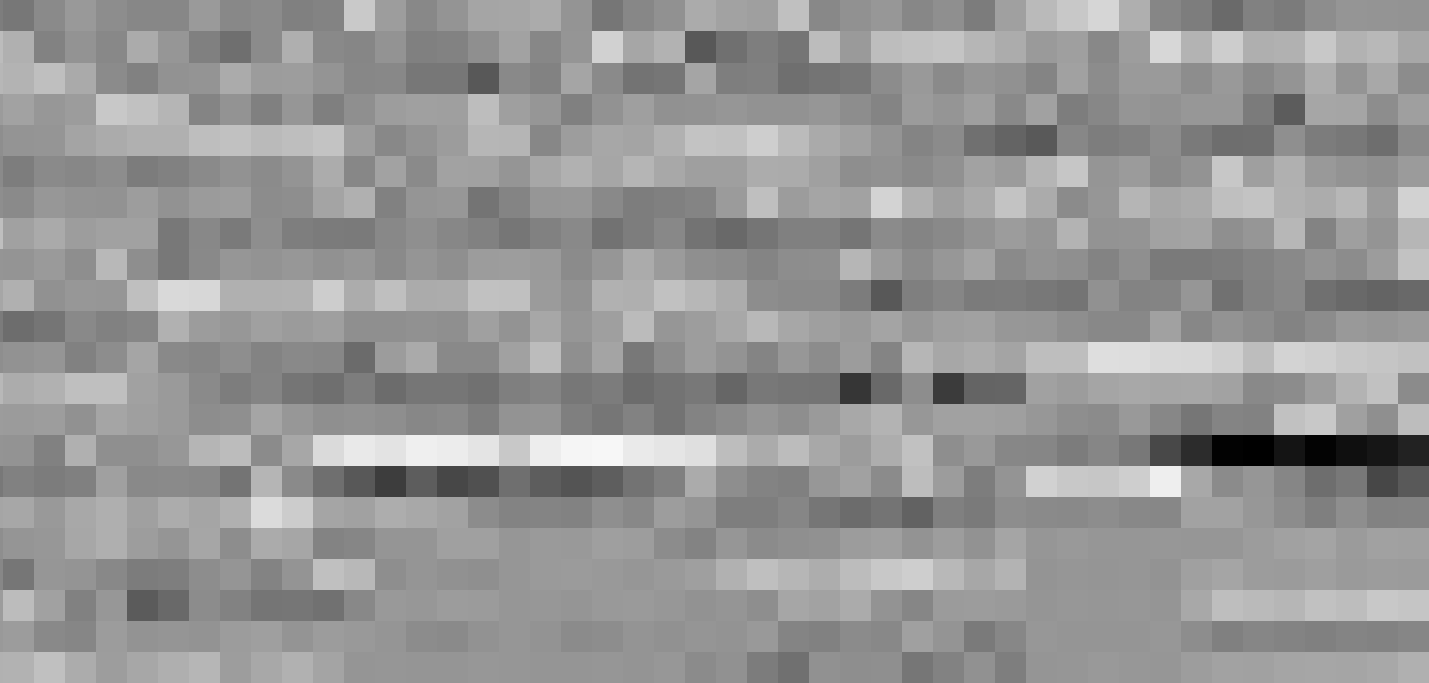}	  	  
	}	
	\subfloat{%
		\includegraphics[width=0.154\textwidth]{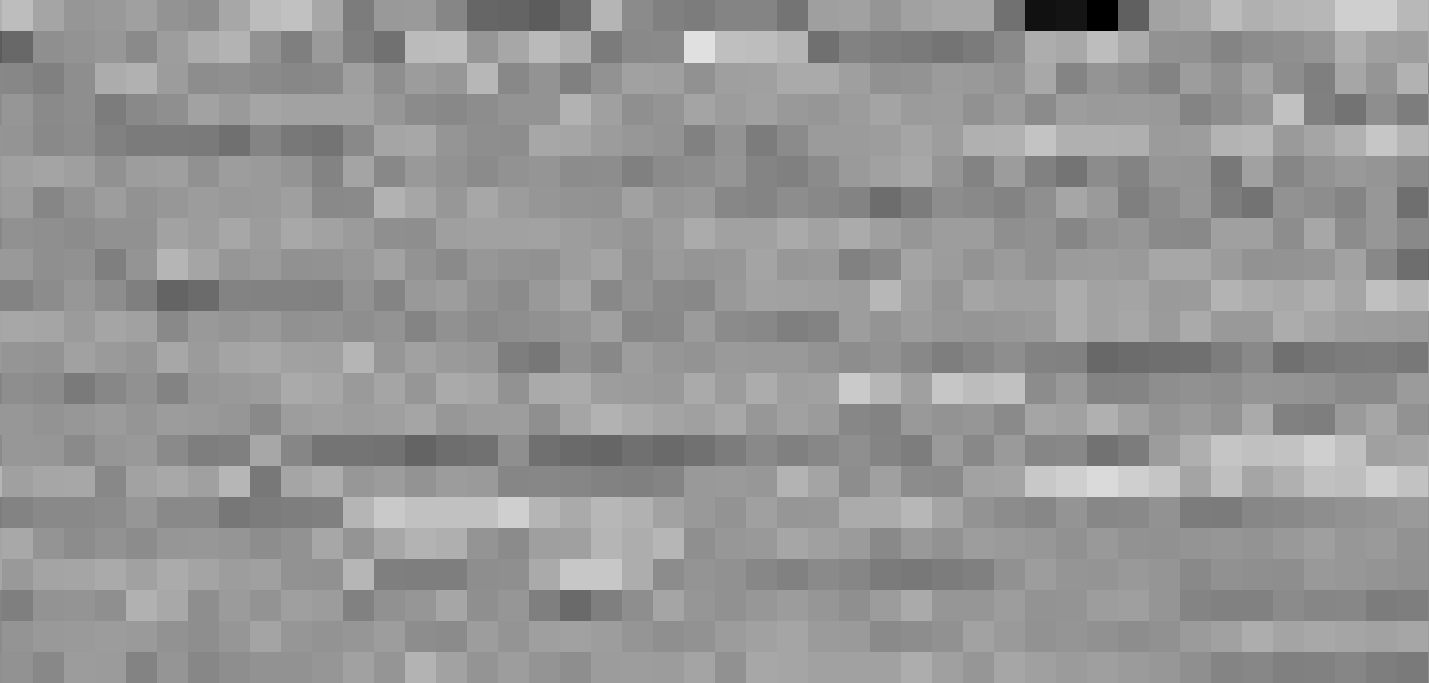}	  
	}
	\subfloat{%
		\includegraphics[width=0.159\textwidth]{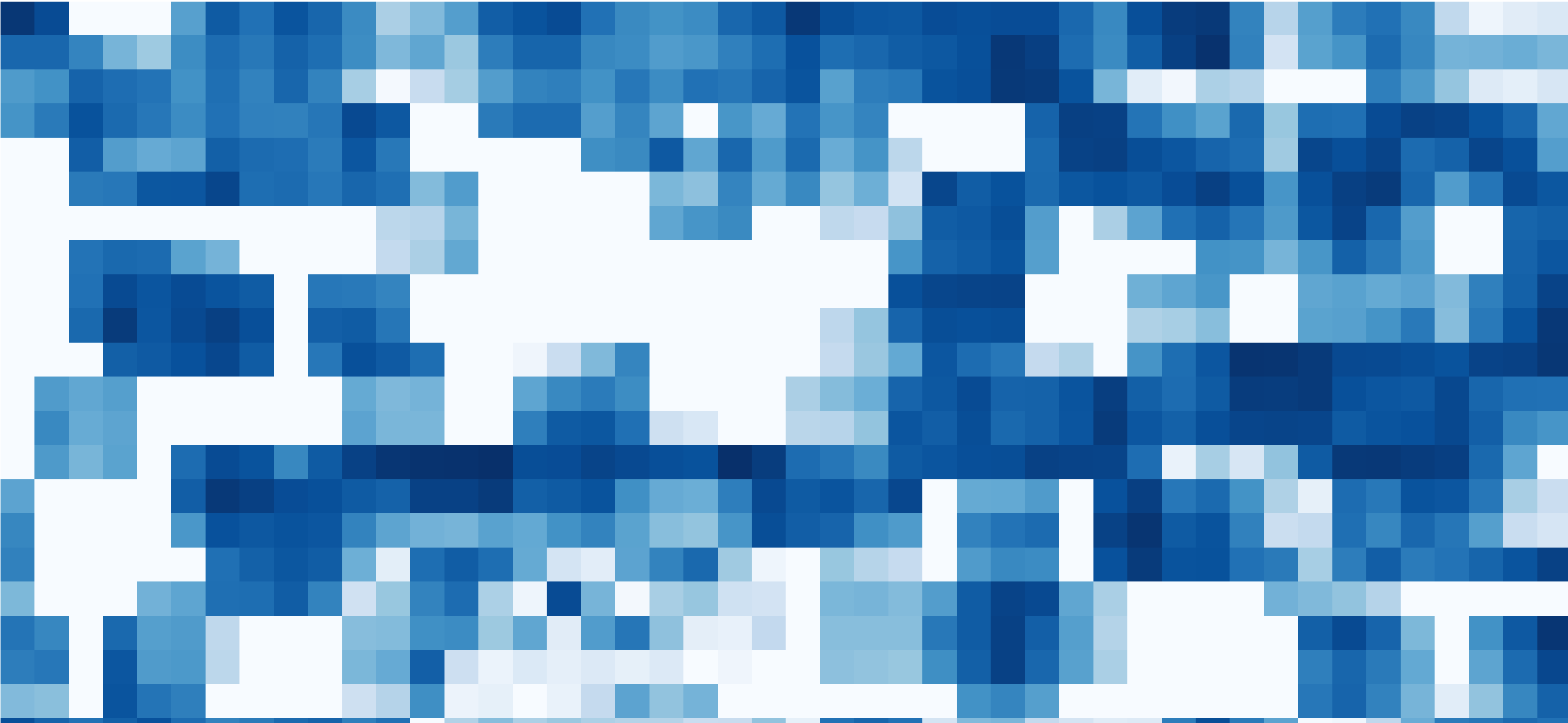}	  
	}  
	\subfloat{%
		\hspace*{-1.2ex} \includegraphics[width=0.017\textwidth]{figures/fig1/bar.png}
	} 	
	\vspace{-2mm}
	\caption{Dependencies between color components in the learned wavelet domain. Dark blue values indicate high correlation computed as centered cosine similarity in a $3 \times 3$ neighborhood. Row 1: Patch of size $H \times W$ in the spatial domain. Row 2-3: Subband HL$_i$ of size $H/2^i \times W/2^i$, $i=2,3$.  }
	\label{fig:intro}
	\vspace{-6mm}
\end{figure}

Therefore, building the compressive autoencoder according to the lifting scheme is an interesting alternative approach, because this way the latent space corresponds to a wavelet decomposition \cite{maiwave++, Ma2020}. This facilitates interpreting the latents and obtaining a meaningful, inherently hierarchical representation. Liu \etal \cite{Liu2021a}, for example, employed a learned wavelet transform to obtain a scalable feature representation both for compression and machine vision tasks.
An end-to-end wavelet compression method called iWave++ \cite{maiwave++} achieves state-of-the art performance for lossy compression. However, the method transforms and codes the color components of an image in the YCbCr color space independently. No correlations between the color channels can therefore be exploited.

In this paper, we address inter-component dependencies in neural wavelet image coding. Fig. \ref{fig:intro} illustrates that the color components Y and Cb have local correlations in the learned wavelet transform domain of an iWave++ model. To exploit these correlations, we propose a novel cross-component context model (CCM) for the neural wavelet compression framework iWave++. In Versatile Video Coding (VVC), three cross-component intra prediction modes use a linear model to predict chroma samples from reconstructed luma samples in the spatial domain \cite{Bross2021}. In contrast, our CCM conditions the entropy model of chroma components on previously coded components in the learned wavelet latent space. 
\vspace{-3mm}
\section{Neural Wavelet Image Coding} \label{sec:iwave++}
\vspace{-3mm}
\noindent
In the following, the end-to-end trainable wavelet coding scheme iWave++ \cite{maiwave++} will be explained. It serves as baseline model for the proposed CCM. In general, neural wavelet image coding supports both lossless and lossy image compression in contrast to other neural image compression approaches.
\vspace{-1mm}
\subsection{Trainable Wavelet Transform}
\vspace{-1mm}
\noindent
To obtain a compressive autoencoder with a wavelet decomposition as latent space, the encoder and decoder are built using the lifting scheme. The lifting scheme allows constructing second generation wavelets \cite{sweldens1995NewPhil}. 
\begin{figure}[tb]
	\centering	
	\subfloat{%
		\includegraphics[width=0.42\textwidth]{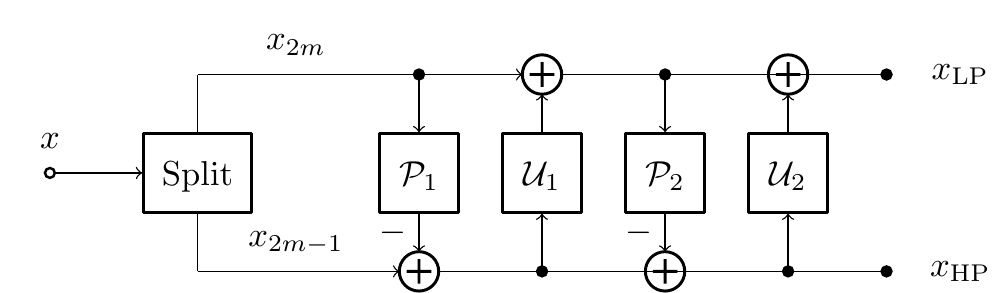}	  
	}
	\caption{1D lifting scheme with two prediction and update filters.}
	\label{fig:fwdLifting}
	\vspace{-2mm}
\end{figure}
The 1D lifting scheme illustrated in Fig.~\ref{fig:fwdLifting} consists of the three stages split, predict, and update. First, the 1D input signal $x$ is split into even $x_{2m}$ and odd samples $x_{2m-1}$. Next, the odd samples are predicted from the even samples. The updated samples $x_{2m-1}=x_{2m-1} - \mathcal{P}_1(x_{2m})$ correspond to a highpass (HP) signal containing less information than $x$. To maintain global properties of $f$, an update step is performed according to $x_{2m} = x_{2m} + \mathcal{U}_1(x_{2m-1})$. After applying two prediction and update filtering steps, the even and odd samples are the lowpass (LP) and HP subband of the 1D Discrete Wavelet Transform (DWT), respectively. The inverse lifting scheme is obtained by reversing the order of the operations and inverting the signs.
To perform a 2D DWT of an input color component, the 1D lifting scheme is first applied to its row dimension and subsequently to the second dimension of the LP and HP subband. Implementing the prediction and update filters by Convolutional Neural Networks (CNNs) gives a trainable DWT. Here, the filters have a residual CNN structure according to Fig.~\ref{fig:filters}. The $3 \times 1$ filter kernels on the skip path are initialized with the CDF 9/7 filter coefficients and tanh is used as nonlinear activation function.
\begin{figure}[tb]
	\centering
	\includegraphics[width=0.3\textwidth]{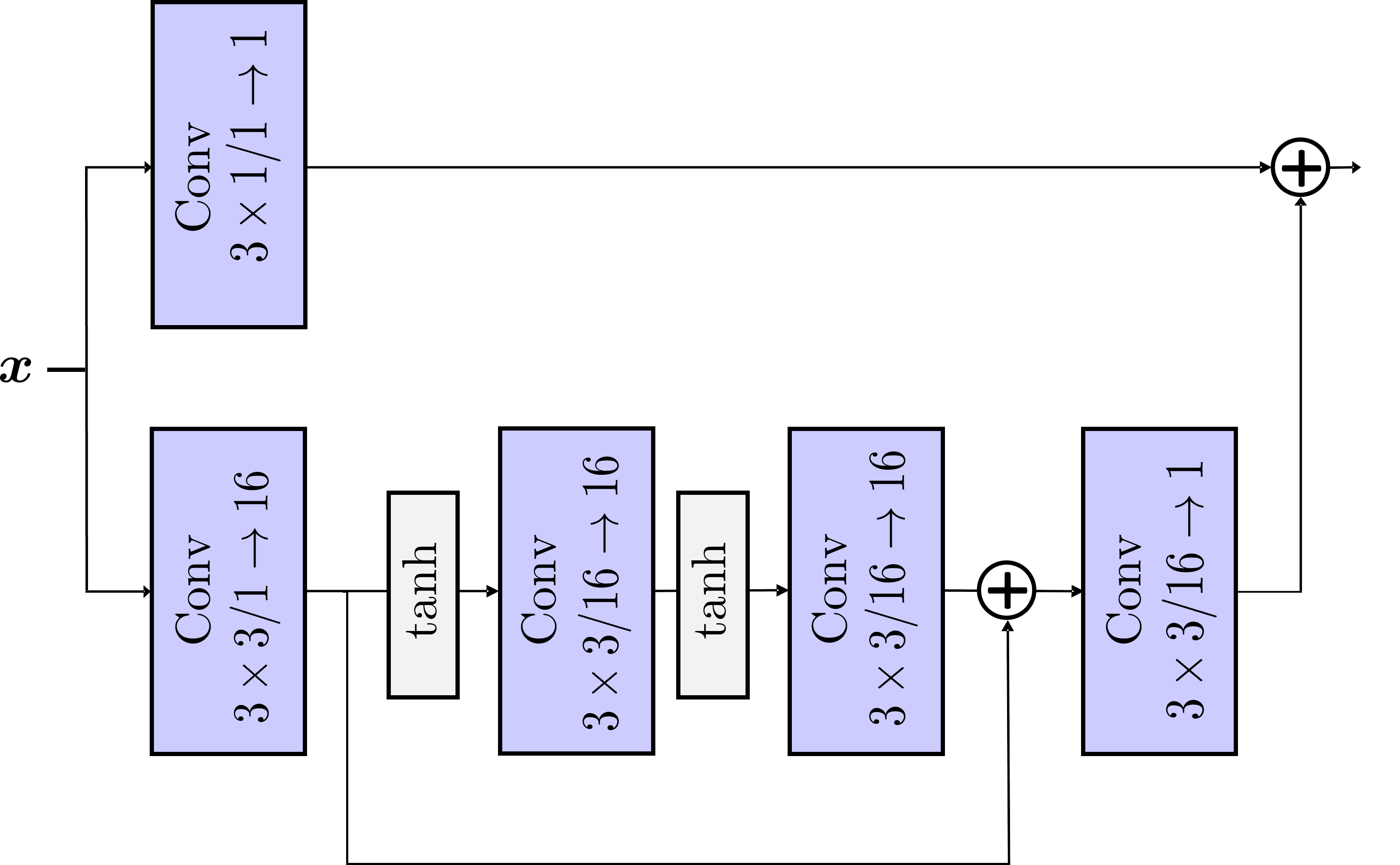}  
	\caption{Implementation of the filters $\mathcal{P}_1, \mathcal{U}_1, \mathcal{P}_2,$ and $\mathcal{U}_2$. The ''Conv'' blocks are convolutional layers, whose parameters are specified as kernel size/input channels $\rightarrow$ output channels. $\bm{x}$ denotes the respective 2D filter input with one channel.}
	\label{fig:filters}
		\vspace{-6mm}
\end{figure}
\vspace{-3mm}
\subsection{End-to-End Trainable Lossy Compression Framework}
\begin{figure}[tb]
	\centering
	\includegraphics[width=0.475\textwidth]{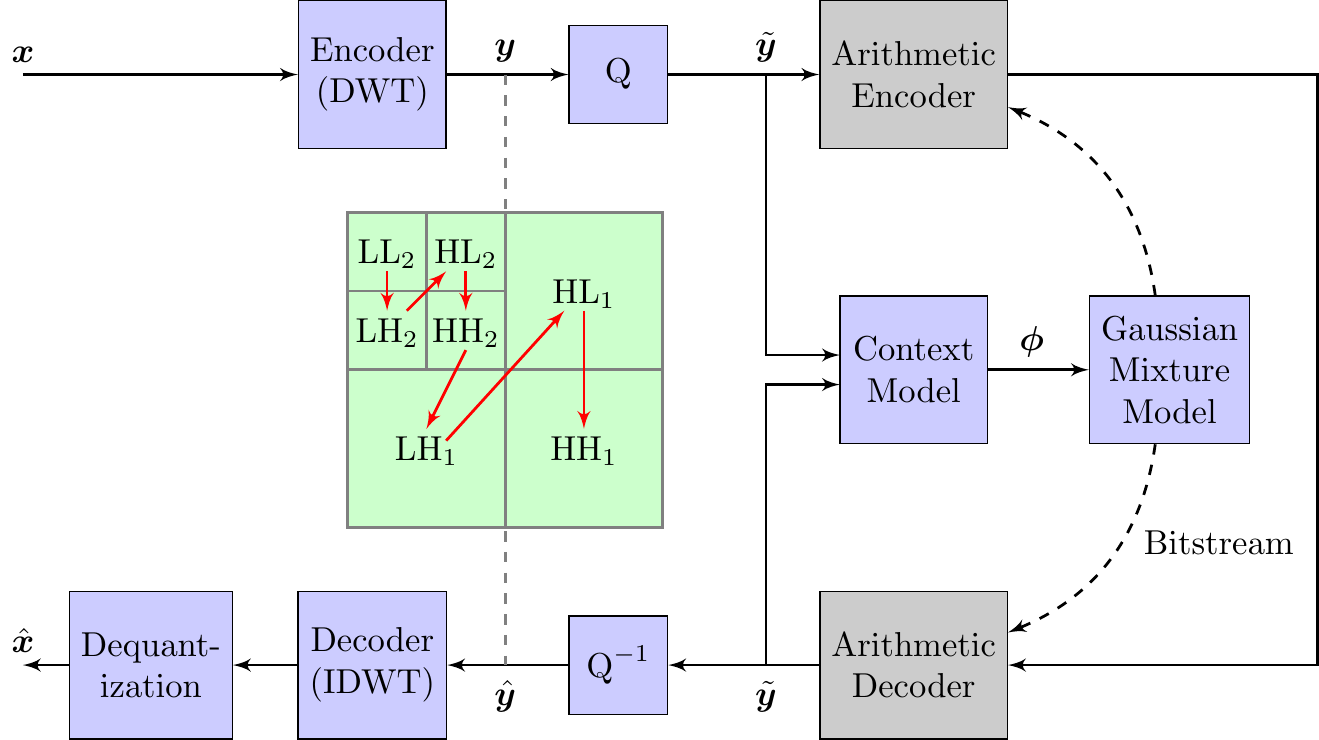}
	\caption{Overview of the baseline model. The red arrows indicate the coding order of the subbands. Trainable modules are shown in blue.}
	\label{fig:overview}
		\vspace{-5mm}
\end{figure}

\vspace{-1mm}
\noindent
Fig.~\ref{fig:overview} provides an overview of the entire image compression framework following iWave++. First, an RGB image to be compressed is converted to the YCbCr color space according to ITU-R BT.601, in which the luma and chroma components have a value range from 0 to 255 for an 8-bit color image \cite{ituycbcr}. The input $\bm{x}$ in Fig.~\ref{fig:overview} is one color component, which is processed independently. The encoder and decoder implement the CNN-based DWT introduced above, i.e., the latent space $\bm{y}$ corresponds to the wavelet transform of $\bm{x}$. Scalar quantization with a trainable parameter $\Delta$ allows optimizing for different rate-distortion tradeoffs. The quantized latents are obtained as $\tilde{\bm{y}} = Q(\bm{y}, \Delta) = \lfloor \bm{y} \cdot \Delta \rceil$. To enable gradient-based optimization, the gradient of the rounding operator $\lfloor \cdot \rceil$ is replaced by an identity function during training. 
The quantized latents $\tilde{\bm{y}}$ are losslessly compressed by adaptive arithmetic coding. The arithmetic encoder and decoder have a shared entropy model, which is a conditional Gaussian Mixture Model (GMM) conditioned on previously coded wavelet coefficients. A neural network-based autoregressive context model estimates the parameters $\bm{\phi}$ for the GMM. Details on the context model will be provided in the subsequent chapter. Note that Ma \etal \cite{maiwave++} use a parameterized factorized entropy model. We use a GMM instead, because its parameters have a meaningful visual interpretation and we did not find that the factorized model performs better than the GMM in our experiments. %
After arithmetic decoding, the encoding process is reversed as shown in Fig.~\ref{fig:overview} in the lower path. Inverse scalar quantization gives the reconstructed latents $\bm{\hat{y}} = \text{Q}^{-1}(\tilde{\bm{y}}, \Delta) = \tilde{\bm{y}} / \Delta $. Next, the decoder applies the inverse DWT (IDWT). Because the decoder shares parameters with the encoder, it cannot take quantization into account in contrast to compressive autoencoders based on the framework by Minnen, Ball\'{e}, and Toderici (MBT2018) \cite{minnen2018joint}. Therefore, a dequantization module consisting of 6 residual CNN blocks (see Fig.~9 in \cite{maiwave++} for details) compensates for quantization artifacts after the IDWT. 
\vspace{-2mm}
\section{Cross-Component Context Model (CCM)} \label{sec:cccm}
\vspace{-1mm}
\noindent
The following chapter introduces the adaptive entropy coding engine that codes the quantized latents based on the entropy parameters predicted by our novel CCM. The wavelet representation of the color components Y, Cb, and Cr is obtained by the same learned transform. A fully autoregressive context model estimates entropy parameters for every wavelet coefficient. The CCM captures dependencies between different components by conditioning the entropy model of the chrominance wavelet coefficients on previously coded subbands.
\vspace{-4mm}
\subsection{Learning Correlations Across Subbands and Components}
\vspace{-1mm}
\setlength{\belowdisplayskip}{6pt} \setlength{\belowdisplayshortskip}{6pt}
\setlength{\abovedisplayskip}{6pt} \setlength{\abovedisplayshortskip}{6pt}
\noindent
Approaches based on the framework MBT2018 \cite{minnen2018joint} achieve a hierarchic entropy model by using a hyperprior, while the wavelet-based approach provides a hierarchic latent space due to its multiresolution analysis. To exploit dependencies between different wavelet decomposition levels and color components, the subbands of each color component are coded sequentially. The coding order of the individual subbands from one color component is visualized in the middle of Fig.~\ref{fig:overview} for two decomposition levels for simplicity. We use $D=4$ decomposition levels. The coding order can be written as $\mathcal{S} = \left\{ \bm{s}_1, \bm{s}_2, \hdots, \bm{s}_{3D+1} \right\}$, where $\bm{s}_1$ corresponds to subband LL$_D$, for example. To extract relevant information for coding the current subband from all previously coded subbands, a context prediction module using Recurrent Neural Networks (RNNs) is employed. The module consists of a sequence of three convolutional Long-Short Term Memory (LSTM) units \cite{convLSTM} computing the predicted context for subband $\tilde{\bm{s}}_{\sidx}$:
\begin{equation}
	g_{\text{RNN}}^{(\text{Y})} ( \tilde{\bm{s}}_{\sidx-1}  )  = \text{LSTM}_1(\text{LSTM}_2(\text{LSTM}_3( \tilde{\bm{s}}_{\sidx-1} ))),\notag
\end{equation}
where each LSTM unit outputs its hidden state with the same spatial dimensions as the input subband $\tilde{\bm{s}}_{\sidx-1}$. The hidden and cell states have 32 feature maps for LSTM$_{1, 2}$ and 3 for LSTM$_{3}$, respectively. The output prediction is one of the three feature maps of LSTM$_{3}$'s hidden state depending on the subband type. When switching from one to the next decomposition level, the hidden and cell states of all LSTM units are upsampled by a factor of 2. $g_{\text{RNN}}^{(\text{Y})} ( \cdot )$ takes the previously coded subband $\tilde{\bm{s}}_{\sidx-1}$ as input, but has access to information from all previously coded subbands $\tilde{\bm{s}}_{\sidx-2}, \hdots, \tilde{\bm{s}}_{1}$ because of the memory capability of the hidden states. 
The long-term context of the first channel to be coded, that is, the Y channel, is obtained as:
\begin{equation}
	\bm{l}_{\sidx}^{(\text{Y})}= 
	\begin{cases}
		g_{\text{RNN}}^{(\text{Y})} ( \tilde{\bm{s}}^{(\text{Y})}_{\sidx-1} ) & \text{if} \enspace  1 < i \leq 4 \\
		\text{cat} \left(   g_{\text{RNN}}^{(\text{Y})} ( \tilde{\bm{s}}^{(\text{Y})}_{\sidx-1} ), \, \text{Up}( \tilde{\bm{s}}^{(\text{Y})}_{\sidx-3} )   \right) & \text{otherwise}, \notag\\
	\end{cases}
\end{equation}
where ''cat'' denotes concatenation along the channel dimension. For the first subband of the Y channel $\tilde{s}^{(\text{Y})}_1$, no context information is available. For the remaining subbands, the RNN-based prediction is combined with a learned convolutional upsampling $\text{Up}(\cdot)$ of the previously coded subband of the same type at a factor of 2.
When coding the subbands of the color components Cb and Cr, the previously coded components are taken into account as well. 
The long-term context of the Cb subbands reads:
\begin{equation}
	\bm{l}_{\sidx}^{(\text{Cb})}= 
	\begin{cases}
		\text{cat} \left( \tilde{\bm{s}}^{(\text{Y})}_{1}, 	g_{\text{RNN}}^{(\text{Y})} ( \tilde{\bm{s}}^{(\text{Y})}_{1} )  \right) & \text{if} \enspace i = 1 \\
		\text{cat} \left( \tilde{\bm{s}}^{(\text{Y})}_{\sidx},   g_{\text{RNN}}^{(\text{C})} ( \tilde{\bm{s}}^{(\text{Cb})}_{\sidx-1} ) \right) &  1 < i \leq 4 \\
		\text{cat} \left( \tilde{\bm{s}}^{(\text{Y})}_{\sidx},  g_{\text{RNN}}^{(\text{C})} ( \tilde{\bm{s}}^{(\text{Cb})}_{\sidx-1} ), \, \text{Up}( \tilde{\bm{s}}^{(\text{Cr})}_{\sidx-3} )   \right) &  i > 4 \notag \\
	\end{cases}
\end{equation}
The subbands of the Y channel and its long-term context serve as additional context information.
The context of the third color component to be coded is obtained as follows:
\begin{equation}
	\bm{l}_{\sidx}^{(\text{Cr})}= 
	\begin{cases}
		\text{cat} \left( \tilde{\bm{s}}^{(\text{Y})}_{1}, 	\tilde{\bm{s}}^{(\text{Cb})}_{1}  \right) & \text{if} \enspace i = 1 \\
		\text{cat} \left( \tilde{\bm{s}}^{(\text{Y})}_{\sidx},  \tilde{\bm{s}}^{(\text{Cb})}_{\sidx},  g_{\text{RNN}}^{(\text{C})} ( \tilde{\bm{s}}^{(\text{Cr})}_{\sidx-1} ) \right) &  1 < i \leq 4 \\
		\text{cat} \left( \tilde{\bm{s}}^{(\text{Y})}_{\sidx},  \tilde{\bm{s}}^{(\text{Cb})}_{\sidx}, g_{\text{RNN}}^{(\text{C})}( \tilde{\bm{s}}^{(\text{Cr})}_{\sidx-1} ), \text{Up}( \tilde{\bm{s}}^{(\text{Cr})}_{\sidx-3} )   \right) &  i > 4.\notag \\
	\end{cases}
\end{equation}
Here, the already coded subbands of Y and Cb are exploited as additional information for coding Cr.
\setlength{\belowdisplayskip}{3pt} \setlength{\belowdisplayshortskip}{3pt}
\setlength{\abovedisplayskip}{3pt} \setlength{\abovedisplayshortskip}{3pt}

\begin{figure}[tb]
				\vspace{-1mm}
	\centering	
	\includegraphics[width=0.49\textwidth]{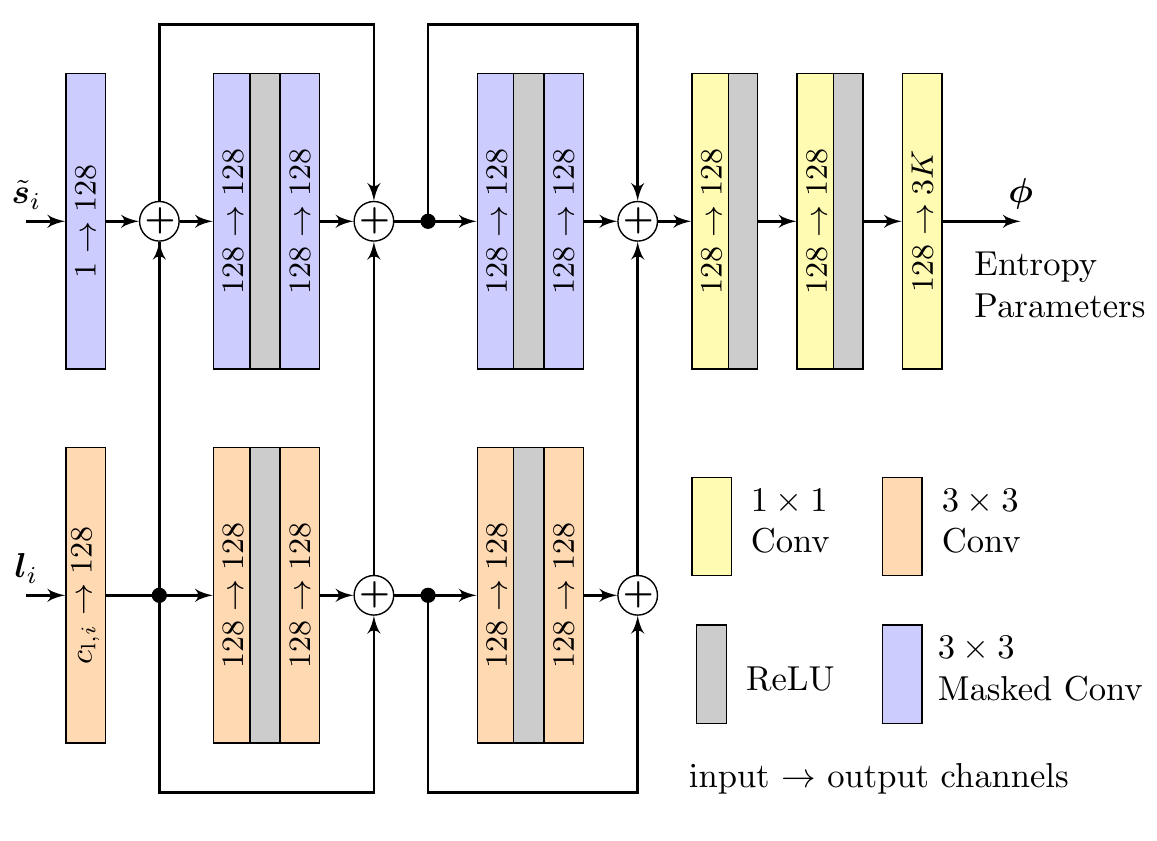}	  
	\vspace{-6mm}
	\caption{Context fusion module $g_{\text{cf}, \sidx}(\cdot)$ for every subband $\tilde{\bm{s}}_i$. Upper path: Autoregressive processing of the current subband. Combined with context information from the previously processed subbands on the lower path. }
	\label{fig:contextFusion}
		\vspace{-5mm}
\end{figure}

A context fusion module $g_{\text{cf}, i}(\cdot)$ predicts the entropy parameters $\bm{\phi}$ for subband $\tilde{\bm{s}}_i$ based on the long-term context $\bm{l}_{\sidx}$ of every color component.
The two chroma components share the RNN-based prediction module $g_{\text{RNN}}^{\text{(C)}}(\cdot)$ and context fusion modules $g_{\text{cf}, i}^{\text{(C)}}(\cdot)$, while the Y channel has dedicated modules $g_{\text{RNN}}^{\text{(Y)}}(\cdot)$ and $g_{\text{cf}, i}^{\text{(Y)}}(\cdot)$. The context fusion modules have the same structure for every color component, which is illustrated in Fig.~\ref{fig:contextFusion}. The lower path processes the long-term context $\bm{l}_i$ of the quantized subband $\tilde{\bm{s}}_i$ to be coded by two residual blocks with $3 \times 3$ convolutions. The number of input channels $c_{\text{l}, i}$ of the first convolution depends on the subband number and color component type. For example, $\bm{l}_{\sidx}^{(\text{Cr})}$ has 4 channels for $i > 4$.
The upper path in Fig.~\ref{fig:contextFusion} processes the subband $\tilde{\bm{s}}_i$ to be coded in line-scan order by autoregressive $3 \times 3$ masked convolutions. Additions on the upper path combine causal information with the long-term context. Finally, a sequence of $1 \times 1$ convolutions reduces the number of feature maps from 128 to $3K$ to obtain the entropy parameters $\bm{\phi}$ for every wavelet coefficient. Overall, the entropy parameter prediction explained above can model correlations within the same subband as well as across subbands and color components.

The conditional probability model for the wavelet coefficient $\tilde{s}_{\sidx, \cidx}$ from subband $i$ at spatial position $j$ is defined as
\begin{align}
	p_{\tilde{s}_{\sidx, \cidx}}(\tilde{s}_{\sidx, \cidx}   | \tilde{\bm{s}}_{\sidx, < \cidx}, \bm{l}_{\sidx}  )   =  & 
	\sum_{k=1}^{K} w_{\sidx, \cidx}^{(k)} \, \mathcal{N}\left(  \mu_{\sidx, \cidx}^{(k)}, \, \sigma_{\sidx, \cidx}^{2(k)}  \right) (\tilde{s}_{\sidx, \cidx})  \notag  \\
	\text{with}  \enspace  \bm{\phi}_{\sidx, \cidx} = & \left\{ \bm{w}_{\sidx, \cidx},  \bm{\mu}_{\sidx, \cidx},  \bm{\sigma}_{\sidx, \cidx}\right\} = g_{\text{cf}, \sidx}(\tilde{\bm{s}}_{\sidx, < \cidx}, \bm{l}_{\sidx} ) \, , \notag
\end{align}
where $\tilde{s}_{\sidx, < \cidx}$ denotes the causal context at position $j$ determined by the masked convolutions of the context fusion module $g_{\text{cf}, i}(\cdot)$. The GMM consists of $K=3$ mixtures, i.e.,  $\bm{\phi}_{\sidx, \cidx}$ contains 9 parameters per coefficient. Each mixture is a Gaussian distribution weighted by the mixing probabilities $\bm{w}_{\sidx, \cidx}$ obtained from a Softmax function.
\vspace{-3mm}
\subsection{Loss Function and Training Strategy}
\vspace{-1mm}
\noindent
The joint probability model $p_{\tilde{\bm{s}}_{\sidx}}(\tilde{\bm{s}}_{\sidx} )$ of subband $\tilde{\bm{s}}_{\sidx}$ is the product of the coefficient probability models $p_{\tilde{s}_{\sidx, \cidx}}(\tilde{s}_{\sidx, \cidx}   | \tilde{\bm{s}}_{\sidx, < \cidx}, \bm{l}_{\sidx}, \bm{\theta}_{\text{cf},i}   )$. It is used to estimate the expected rate $R$ during training, which is lower bounded by the entropy of the wavelet coefficients. The rate-distortion loss $\mathcal{L} = R + \lambda D$ for one color component
$\bm{x}$ is thus computed as: 
\begin{equation}
	\mathbb{E}_{\bm{x} \sim p_{\bm{x}}}     \left[     \sum_{\sidx=1}^{3D+1} \sum_{\cidx=1}^{J} - \log_2 ( p_{\tilde{\bm{s}}_{\sidx, \cidx}} (\tilde{\bm{s}}_{\sidx, \cidx}))      \right]     + \lambda \cdot  \mathbb{E}_{\bm{x} \sim p_{\bm{x}}}  ||\bm{x} - \bm{\hat{x}}||_2^2, \notag
\end{equation}
with the mean squared error as distortion measurement. $J$ is the number of wavelet coefficients in subband $\tilde{\bm{s}}_{\sidx}$. The unknown distribution of natural images $p_{\bm{x}}$ is approximated by averaging over the elements of a mini-batch during training. First, the entire model is trained with luminance data only, i.e., $\bm{x}$ is the Y component of an input image and $\mathcal{L}^{(\text{train})} = \mathcal{L}^{(\text{Y})}$. Subsequently, the parameters of the wavelet transform, dequantization module, and the quantization parameter $\Delta$ are kept fixed and only the parameters of the chroma context fusion and prediction modules are trained using $\mathcal{L}^{(\text{train})}  = \frac{1}{2} (\mathcal{L}^{(\text{Cb})} + \mathcal{L}^{(\text{Cr})} )$. The parameters of the chroma RNN prediction and context fusion modules are initialized based on the respective luma modules. Additionally, the cell and hidden states of the chroma context prediction module $g_{\text{RNN}}^{\text{(C)}}(\cdot)$ are adopted from the respective states obtained from processing the Y channel of every mini-batch element.
\vspace{-1mm}
\section{Experiments and Results} \label{sec:experiments}
\vspace{-1mm}
\subsection{Experimental Setup}
\vspace{-1mm}
\noindent
We implemented the network as shown and explained above using the PyTorch framework. We use the model trained on luminance data only as baseline model, which we refer to as ''iWave++ Y'' \cite{maiwave++}. This model achieves comparable results to the original iWave++ models published by the authors, for which no code for bitstream generation and training is available. Therefore, we use our own implementation and training configuration.
We train on the 800 DIV2K \cite{div2k1} and 585 CLIC2020 professional training images \cite{CLIC2020} as well as as the Vimeo90K triplet data set \cite{xue2019video}. We use a batch size of 16 and randomly crop $256 \times 256$ patches from the training images. We train 6 models by choosing the rate-distortion tradeoff parameter according to $\lambda =  \left\{ 0.003, 0.007, 0.01, 0.03, 0.05, 0.08 \right\}$. For the rate point $\lambda = 0.08$, we train the baseline iWave++ Y model for 27 epochs on the luma training data. By finetuning this model for 3 epochs, we obtain the remaining rate points. The CCM module is trained for 9 epochs based on the obtained iWave++ Y models. We refer to the cross-component model as ''iWave++ CCM''. The optimizer is AdamW \cite{loshchilov2018decoupled} with an initial learning rate of $\num{1e-04}$. We use a cosine learning rate schedule with a minimum learning rate of $\num{1e-06}$. 
For testing, we use the Kodak and Tecnick \cite{Asuni2013} data set. We compare our method to the compression standards JPEG2000, the HM 16.24 and VTM 14.0 test model in all intra configuration as well as the learning-based methods MBT2018 \cite{minnen2018joint}, ICLR19 \cite{iclr2019}, TPAMI21 \cite{Hu2021} and ANFIC \cite{Ho2021b}. 
\vspace{-3mm}
\subsection{Experimental Results}
\begin{figure}[tb]
	\includegraphics[width=0.38\textwidth]{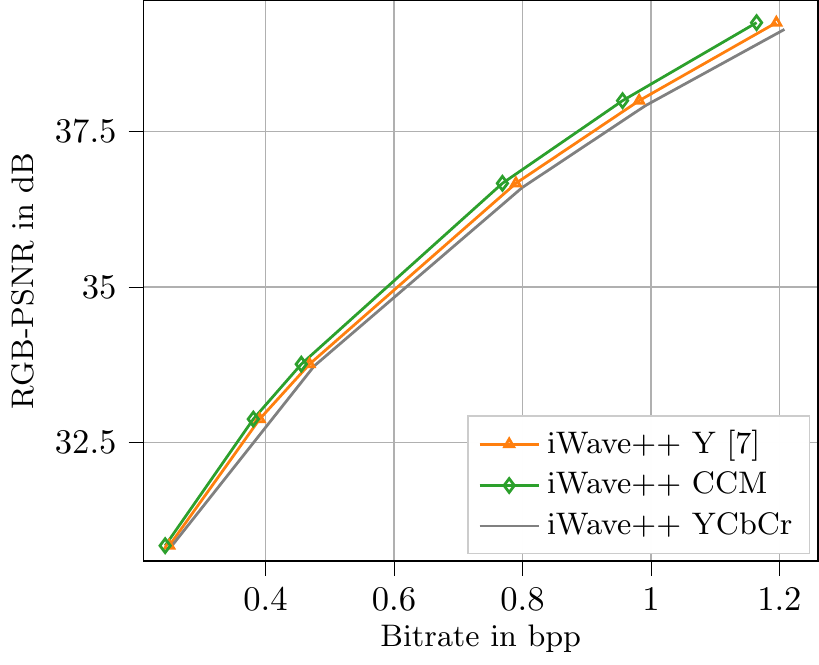}
	\vspace{-5mm}
	\caption{Rate-distortion evaluation on the Kodak data set.}
	\label{fig:kodak1}
\vspace{-4mm}
\end{figure}

\vspace{-1.5mm}
\begin{table}[tb]
	\vspace{-5mm}
	\captionsetup[table]{skip=0pt} 
	\caption{Rate-distortion evaluation on Kodak and Tecnick. }
	\vspace{0.5mm}
	\renewcommand{\arraystretch}{1.2}
	\resizebox{0.5\textwidth}{!}{
		\begin{tabular}{l|l|l|l|l|l}
			\multicolumn{3}{l|}{(a) \textbf{BD rate anchor iWave++ Y \cite{maiwave++}}}& \multicolumn{3}{l}{(b) \textbf{BD rate anchor HM}} \\
			\hline
			& Kodak & Tecnick  &
			& Kodak & Tecnick  \\
			\hline
			iWave++ &  \multirow{2}{*}{+2.31~\%}&  \multirow{2}{*}{+6.27~\%}  &
			MBT2018\cite{minnen2018joint} & -11.28~\%& -17.59~\% \\
			YCbCr & &   &
			ICLR2019\cite{iclr2019} & -11.43~\% & -15.09~\% \\
			
			iWave++ CCM &   \textbf{-2.63~\%} & \textbf{-1.65~\%}&   TPAMI21\cite{Hu2021} & -16.16~\% & -19.95~\% \\

		\multicolumn{3}{l|}{} & ANFIC\cite{Ho2021b} & -21.70~\% & \textbf{-26.67~\%} \\
		\multicolumn{3}{l|}{} & iWave++ CCM & \textbf{-23.23~\%} & -20.51~\%\\
		\end{tabular}
	}\vspace{-5mm}
	\label{tab:tecnick}
\end{table}
\noindent
First, we compare our novel iWave++ CCM to the baseline iWave++ Y. As can be seen from the rate-distortion evaluation Tab.~\ref{tab:tecnick}(a), 
iWave++ CCM achieves Bj{\o}ntegaard delta (BD) rate savings \cite{bdrate} of 2.6~\% and 1.6~\% compared to iWave++ Y on Kodak and Tecnick, respectively. The chroma components account for about 19~\% of the bits used to compress the Kodak images with the highest quality iWave++ Y model and for about 16~\% with the lowest quality model. This indicates that the chroma components are more sparse than the luma component at lower bitrates. The gains achieved by the proposed CCM are therefore larger at higher bitrates as can be seen in Fig.~\ref{fig:kodak1}, at which more correlations between the components can be exploited. For comparison, ''iWave++ YCbCr'' is also evaluated on the test data sets. Here, we train three individual models independently on every color component. With iWave++ YCbCr, the BD rate deteriorates by about 2.3~\% on Kodak and 6.2~\% on Tecnick compared to iWave++ Y, which applies one model trained on the luma channel to all three components. This suggests that it is harder to learn a good wavelet representation from the chroma components containing less information than the luma data. Finetuning a dedicated wavelet transform jointly with the CCM resulted in slightly worse performance in our experiments.
\begin{figure}[tb]
	\includegraphics[width=0.4\textwidth]{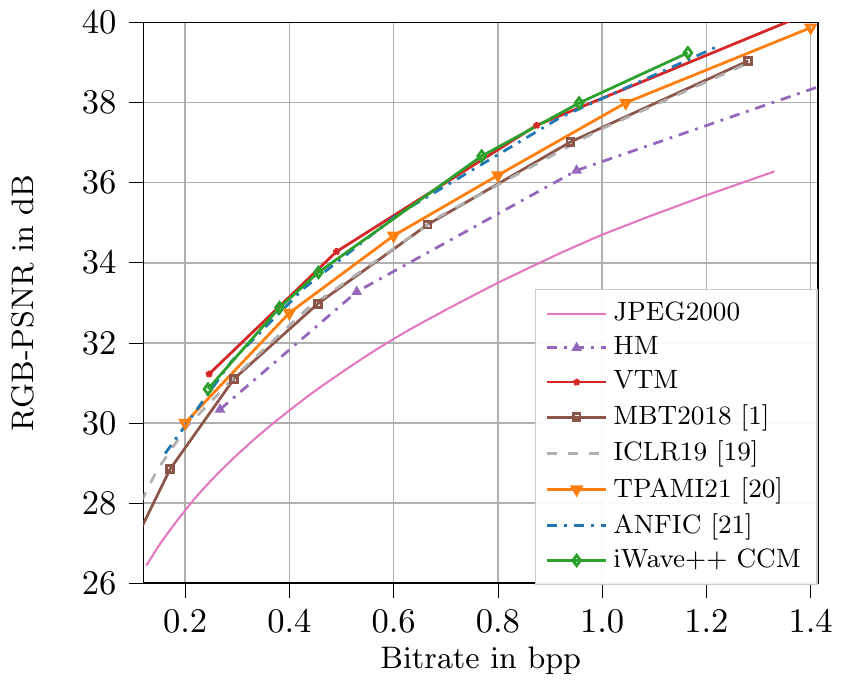}
		\vspace{-5mm}
	\caption{Rate-distortion evaluation on the Kodak data set.}
	\label{fig:kodak2}
	\vspace{-5mm}
\end{figure}

The curves in Fig.~\ref{fig:kodak2} and BD rate savings Tab.~\ref{tab:tecnick}(b) compare our iWave++ CCM to other learning-based and conventional compression methods with HM as anchor. On the Kodak data set, our method achieves a BD rate saving of 23.23~\% outperforming ANFIC by 1.53~\%, while the highest rate saving of 25.05~\% is achieved by VTM. Note that VTM performs better at lower bitrates compared to the learning-based approaches, while the latter outperform VTM at higher bitrates (see Fig.~\ref{fig:kodak2}). This behavior is similar on the Tecnick data set, on which ANFIC achieves the highest BD rate saving of 26.67~\% followed by VTM with a saving of 23.10~\%. The performance of our method is in between TPAMI21 and VTM and thus competitive with state-of-the art compression techniques. 

To assess computational complexity, we measure the average decoding time for the Kodak images on a NVIDIA Quadro RTX 8000 GPU in seconds. Our iWave++ CCM is the slowest method with 3084s on average. ANFIC \cite{Ho2021b} takes 438s, TPAMI21 \cite{Hu2021} 3s, ICLR19 \cite{iclr2019} 77s, and MBT2018 \cite{minnen2018joint} 12s. 
Note that our CCM only accounts for a complexity increase of about 3\% relative to the baseline model. To reduce the runtime of the baseline model, block parallel coding \cite{maiwave++} or a parallelized spatial context model \cite{He2021} can be employed.
\vspace{-4mm}
\section{Conclusion}\label{sec:conclusion}
\vspace{-2mm}
\noindent
In this paper, we introduced a novel cross-component context model called CCM for the neural wavelet image coding framework iWave++ \cite{maiwave++} and provided a matching training strategy as well. By exploiting correlations between the learned wavelet representation of the luma and chroma components, our method achieves BD rate savings of 2.6~\% and 1.6~\% on the Kodak and Tecnick data set, respectively. Our method is competitive with VVC and state-of-the-art learned image compression approaches, while being interpretable. In future work, we will examine the extension of our method to video coding as a promising end-to-end trainable approach supporting both lossless and lossy video compression.  

\bibliographystyle{IEEEbib}
\bibliography{wavelets}

\end{document}